\documentclass[twocolumn]{aastex631}

\shorttitle{Breathing pulses and ZZ Ceti pulsations}
\shortauthors{C\'orsico \& Althaus}

\usepackage{longtable}
\usepackage[flushleft]{threeparttable}
\usepackage{tabularx}
\usepackage{multirow}
\usepackage{graphicx}
\usepackage{amsmath,amssymb}
\usepackage{color}
\usepackage{units}
\usepackage{epstopdf}
\usepackage{hyperref}
\usepackage{multirow}
\usepackage{url}
\usepackage{subfigure}
\usepackage{rotating}
\usepackage{enumitem}\setlist[description]{font=\textendash\enskip\scshape\bfseries}
\usepackage{booktabs}

\newcommand{\beq}{\begin{equation}}
\newcommand{\eeq}{\end{equation}}
\newcommand{\bdm}{\begin{displaymath}}
\newcommand{\edm}{\end{displaymath}}
\newcommand{\T}[1]{\tilde{#1}}

\definecolor{Gray}{gray}{0.9}
\definecolor{orange}{rgb}{0.9,0.5,0}

\newcommand{\ztfink}[1]{ZTF-Fink}

\begin{document}

\title{The impact of breathing pulses during core-helium burning on the core 
  chemical structure and pulsations of hydrogen-rich atmosphere white dwarfs}

\author[0000-0002-0006-9900]{Alejandro H. C\'orsico}
\affil{Grupo de Evoluci\'on Estelar y Pulsaciones. Facultad de 
           Ciencias Astron\'omicas y Geof\'{\i}sicas, 
           Universidad Nacional de La Plata, 
           Paseo del Bosque s/n, 
           (1900) La Plata, 
           Argentina}
\affil{Instituto de Astrof\'{\i}sica de La Plata, 
           IALP (CCT La Plata), 
           CONICET-UNLP}

\author[0000-0003-0658-0459]{Leandro G. Althaus}
\affil{Grupo de Evoluci\'on Estelar y Pulsaciones. Facultad de 
           Ciencias Astron\'omicas y Geof\'{\i}sicas, 
           Universidad Nacional de La Plata, 
           Paseo del Bosque s/n, 
           (1900) La Plata, 
           Argentina}
\affil{Instituto de Astrof\'{\i}sica de La Plata, 
           IALP (CCT La Plata), 
           CONICET-UNLP}

\email{acorsico@fcaglp.unlp.edu.ar}

\begin{abstract}

Breathing pulses are mixing episodes that could develop during the core-helium burning phase of low- and intermediate-mass stars. The occurrence of breathing pulses is expected to bear consequences on the formation and evolution of white dwarfs, particularly on the core chemical structure, which can be probed by asteroseismology. We aim to explore the consequences of breathing pulses on the chemical profiles and pulsational properties of variable white-dwarf stars with hydrogen-rich envelopes, known as ZZ Ceti stars. We compute stellar models with masses of $1.0 M_{\odot}$  and $2.5 M_{\odot}$ in the zero-age main sequence, and evolve them through the core-helium burning phase to the thermal pulses on the asymptotic giant branch, and finally to advanced stages of white-dwarf cooling. We compare the chemical structure of the core of white dwarfs whose progenitors have experienced breathing pulses during the core-helium burning phase with the case in which breathing pulses have not occurred. We find that, when breathing pulses occur, the white-dwarf cores are larger and the central abundances of oxygen are higher than for the case in which the breathing pulses are suppressed, in line with previous studies. However, the occurrence of breathing pulses is not sufficient to explain the large cores and the excessive oxygen abundances that characterize recently derived asteroseismological models of pulsating white dwarfs. We find absolute differences of up to $\sim 30$ seconds when we compare pulsation periods of white dwarfs coming from progenitors that have experienced breathing pulses with the case in which the progenitors have not suffered breathing pulses.
 \end{abstract}


\section{Introduction}

White-dwarf (WD) stars constitute the final product of the evolution of low- and intermediate-mass stars, that is, stars with initial masses lower than $\sim 10-11 M_{\sun}$, depending on their initial metallicity \citep{2015ApJ...810...34W}. Once low- and intermediate-mass stars leave the Main Sequence, when the central hydrogen (H) content has been exhausted, helium (He) ignites in the central regions, giving rise to the core-He burning (CHeB) phase. During  the CHeB stage, the structure of the core of stars is supposed to consist roughly of a central He-burning convection zone that is surrounded by a He-rich region that is not convective. The structure of the edge of the convective core is highly uncertain and has a direct impact on the duration of this phase and the pulsational properties of stars in the CHeB stage (e.g., subdwarf B stars, horizontal branch stars, RR Lyrae stars, and red giant stars), and also on the subsequent evolution. In particular, the treatment of the uncertainties in the mixing of material at the edge of the convective core during the CHeB phase leads to different possible chemical structures of the core of the emerging  WDs. 

One of the most uncertain aspects of the CHeB phase ---that has been (and continues to be) subject of debate--- is the possible occurrence of mixing episodes called "breathing pulses" \citep[BPs;][]{1973ASSL...36..221S,1985ApJ...296..204C}  that take place late in the CHeB phase. At this stage, the position of the formal convective boundary becomes unstable to mixing episodes and BPs may arise as a rapid growth in the mass of the convective core when the central He abundance is very low  ($X_{\rm He} \lesssim 0.10$) and the $^{12}$C($\alpha,\gamma)^{16}$O reaction dominates 
on the triple-$\alpha$ reaction. The effect of BPs is to carry fresh He from the non convective mantle into the convective core,  thus prolonging the CHeB lifetime. As a result of BPs,  the mass of the convective core grows significantly and the central abundance of oxygen ($^{16}$O) increases  \citep{2016MNRAS.456.3866C,2017MNRAS.472.4900C}. Hence, the occurrence or not of BPs in low- and intermediate-mass stars is expected to bear consequences on the formation and evolution of WDs, in particular their core chemical structure. BPs should not be 
confused with the "thermal pulses" (TPs) that take place late in the AGB evolution. They are distinct phenomena in the context of stellar evolution. TPs typically refer to episodic events during the asymptotic giant branch (AGB) phase, characterized by a sudden increase in the rate of He burning in the shell surrounding the stellar core. This results in a temporary expansion of the outer envelope and an increase in luminosity \citep[see, for instance,][]{2013sse..book.....K}.

The structure and chemical constituents of the inner cores of WDs ---besides other important properties such as stellar mass, rotation, etc--- can be assessed through asteroseismology, a modern approach based on the comparison of the observed oscillation periods of pulsating WDs with theoretical periods calculated on appropriate stellar models 
\citep{2008ARA&A..46..157W, 2008PASP..120.1043F, 2010A&ARv..18..471A, 2019A&ARv..27....7C}. 
Recent asteroseismological studies of pulsating DA (H-rich atmospheres) WDs ---also called DAV or ZZ Ceti stars--- and pulsating DB (He-rich atmospheres) WDs ---also known as DBV or V777 Her stars--- based on static parametric models of WDs have been carried out by \cite{2018Natur.554...73G, 2021arXiv210615701G} and \cite{2021arXiv210703797C}. In the case of four DAV stars  (SDSS~J1136+0409, EPIC~220347759, KIC~11911480, and L~19$-$2),  and a DBV star (KIC~08626021), these studies have resulted in asteroseismic models characterized by cores and central O abundances that are substantially larger than those predicted by canonical evolutionary calculations  of WD progenitors \citep{2010ApJ...716.1241S, 2010ApJ...717..183R,2010ApJ...717..897A, 2022A&A...663A.167A}. 
In this sense, \cite{2018ApJ...867L..30T} have shown that the inclusion of neutrino cooling in the models of DB WDs could lead to a different asteroseismological model for KIC~08626021 than the one found by \cite{2018Natur.554...73G}. However, the inclusion of this effect by \cite{2019A&A...628L...2C} seems to lead to a seismological model for this star that is qualitatively similar to that originally found by \cite{2018Natur.554...73G}. \cite{2019A&A...630A.100D} explored in detail the possible changes in the C$^{12}(\alpha, \gamma)^{16}$O reaction rate, screening processes, microscopic diffusion, and overshooting efficiency during 
core-He burning that  could lead to a chemical structure similar to that found by \cite{2018Natur.554...73G} for a DBV star such as KIC~08626021 through asteroseismology. They found that within the current understanding of WD formation from single-star evolution, it is virtually impossible to reproduce the most important asteroseismologically derived features of the chemical structure of KIC~08626021. An additional criticism of \cite{2018Natur.554...73G} results comes from the analysis of \cite{2022RNAAS...6..244B}, who has shown that the asteroseismic radius determination reported by these authors for KIC~08626021 is $6\sigma$ discrepant with constraints from {\it Gaia} astrometry,  calling into question the other results of that asteroseismic analysis, especially the high (central) O abundance that stellar evolutionary models are not able to reproduce. 

 The larger cores and central O abundances reported by \cite{2018Natur.554...73G, 2021arXiv210615701G} and \cite{2021arXiv210703797C}  have raised the question of whether it is possible that such chemical structures of the WD cores may  certainly be the result of physical processes during the evolution of the progenitor stars. Recently, \cite{2022FrASS...9.9045G} have proposed a plausible explanation for the large cores with high central O abundances characterizing  the asteroseismological models of some pulsating WD stars, as being due to the fact that BPs do take place during the CHeB phase of the WD progenitors.  However, the occurrence of BP during CHeB phase is  at odds with the observed ratio of asymptotic giant branch to horizontal branch stars in globular clusters and with asteroseismological inferences in red clump stars in the Kepler field  \citep{2016MNRAS.456.3866C,2017MNRAS.472.4900C}. Given the intriguing suggestion proposed by \cite{2022FrASS...9.9045G}, which, at the same time, generates controversy concerning star counts in globular clusters and the peak of the luminosity distribution function, a thorough exploration of this concept is warranted.

In this work, we examine the chemical structure of the core of WDs whose progenitors have experienced BPs during the CHeB phase, and compare it to the case in which BPs have not occurred.  We will specifically concentrate on the case of DA WDs; however, the conclusions drawn from our analysis also hold true for DB WDs resulting from isolated evolution. We explore the mass range of interest for ZZ Ceti stars. In addition, we study the impact of the occurrence of BPs on the $g$-mode period spectrum of WD models representative of 
ZZ Ceti stars. The paper is organized as follows. In Sect. \ref{evolutionary} we  
describe the evolutionary computations of WDs from the ZAMS taking into account and neglecting BPs. In Sect.  \ref{chemical_profiles} we analyze the chemical structure of our resulting WD models. 
We devote Sect. \ref{pulsation} to describing our analysis of the impact of core BPs experienced by WD progenitors on the period spectrum of ZZ Ceti models. Finally, in Sect. \ref{conclusions} we 
summarize our findings.

\begin{figure}
\begin{center}
\includegraphics[clip,width=1.0\columnwidth]{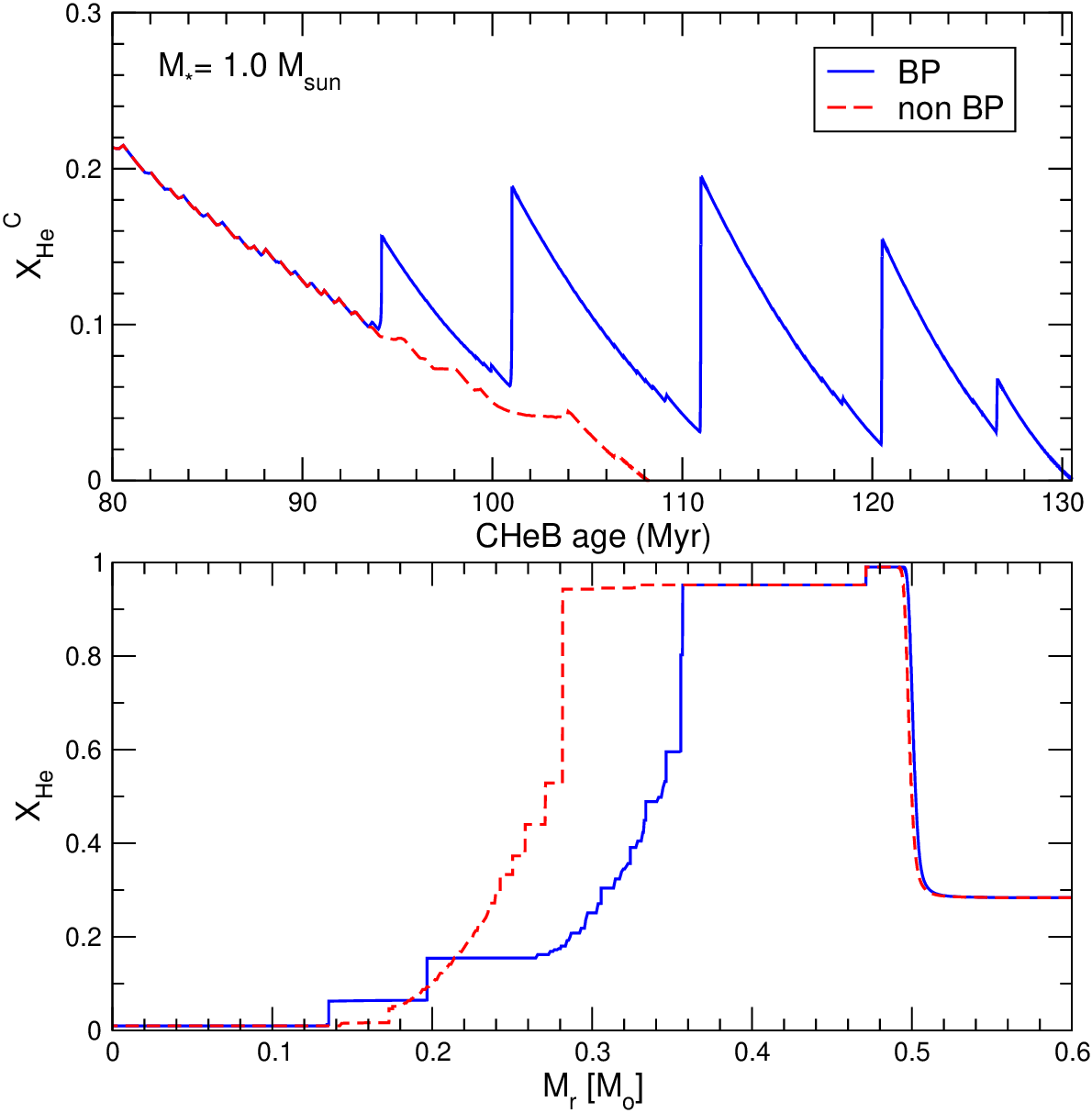}
        \caption{Upper panel: core He abundance (by mass) in terms of the CHeB age. Solid  (dashed) line  corresponds to the case in which BPs during the CHeB phase have been allowed (suppressed). Bottom panel: inner He abundance distribution at the end of CHeB phase when the central He abundance is $X_{\rm He} \approx 0.01$ for the two situations illustrated in the upper panel.}
        \label{fig:hecore}
\end{center}
\end{figure}

\section{Evolutionary computations and treatment of breathing pulses}  
\label{evolutionary}  

We have considered two sets of evolutionary sequences. In one of
this set, the occurrence of BPs during the CHeB phase has been considered ("BP case"). In the other set, BPs have been neglected ("non-BP case").
All of the sequences were computed for metallicity $Z=0.01$ starting from the ZAMS and evolved through the CHeB phase, to the thermal pulses  on the asymptotic giant branch (AGB) and finally to the domain of the ZZ Ceti stars at advanced stages of WD cooling. 
For this purpose, we have used the stellar evolution  code {\tt LPCODE}, developed by the La Plata group \citep{2005A&A...435..631A, 2013A&A...555A..96S, 2015A&A...576A...9A, 2016A&A...588A..25M,2022A&A...663A.167A}. Our treatment for the progenitor evolution considers new observational constraints and recent advances in the micro-  and macro-physics involved in the modeling of core He burning, AGB,  and thermally pulsing phase  \cite[see][for details]{2016A&A...588A..25M}. We focus our study on average-mass ($0.53 \lesssim M_{\star}/M_{\odot} \lesssim 0.60$) pulsating WDs resulting from the single evolution of low-mass progenitors. In particular, we consider WD progenitors of initial mass $M_{\star}= 1.0$ and $2.5 M_{\odot}$. 

\begin{table*}
\caption{Basic  model  properties   for our sequences.}
\centering
\begin{tabular}{lcccccccc}
\hline
\hline
$M_{\rm ZAMS}\, (M_{\sun})$ &$M_{\rm WD}\, (M_{\sun})$ & $\log M_{\rm H}\, (M_{\sun})$ & $t_{\rm CHeB}$~(Myr) & 
$X_{\rm ^{16}O}$ & $N_{\rm TP}$ & $N_{\rm BP}$ & C/O\\
\hline
\multicolumn{8}{c}{BP}\\
1.0  & 0.5363 & $-3.612$ & 130.4 & 0.808 & 4 & 5 & 0.29 \\
2.5  & 0.5903 & $-4.149$ &  177.3 &  0.804 &  12 & 3 & 2.31 \\
\hline
\multicolumn{8}{c}{NON-BP}\\
1.0  & 0.5343 & $-3.595$ & 108.2 & 0.724 & 4 & 0  & 0.29 \\
2.5  & 0.5868 & $-4.147$ &  157.1 &  0.749 &  13 & 0 & 2.48\\
\hline
\end{tabular}

$M_{\rm  ZAMS}$: initial  mass, $M_{\rm  WD}$: WD mass, $\log  M_{\rm H}$:
logarithm of  the mass of H left in the star at the maximum
  effective  temperature  at  the  beginning of  the  cooling  branch, $t_{\rm CHeB}$: lifetime during CHeB phase, $X_{\rm ^{16}O}$: central oxygen abundance, $N_{\rm TP}$:  number of  thermal pulses,  $N_{\rm BP}$:  number of BP pulses, C/O: surface carbon ($^{12}$C) to $^{16}$O ratio after departure from the AGB.
\label{tabla1}
\end{table*}

The evolutionary history of progenitor star determines
the internal chemical profile of the WDs, and  thus their pulsational properties. This is particularly relevant concerning the evolutionary stages corresponding to the CHeB phase.  In order for our models  to experience core BPs during this phase we closely follow the mixing scheme presented in \cite{2016MNRAS.456.3866C,2017MNRAS.472.4900C}. To this end, we adopt their standard-overshoot  model which naturally leads to the occurrence of BPs. Here, time-dependent overshoot mixing is implemented with an exponential decay in the diffusion coefficient $D_{\rm OV}$ beyond all convective boundary according to  $D_{\rm OV}= D_{\rm C}\exp(-2z/H_{\rm v})$ where $D_{\rm C}$ is the diffusion coefficient at the edge of  the convection zone, $z$ is the  radial distance from the
boundary of the  convection zone, $H_{\rm v}= f  H_{\rm P}$, where the free parameter $f$ is a measure of the extent of the overshoot region, and  $H_{\rm  P}$ is  the  pressure  scale  height at  the  convective boundary.  In  this study, we  have adopted $f= 0.005$ for the CHeB phase. At this value, our code predicts the maximum CHeB lifetimes and central oxygen abundances. 
Non-BP sequences were computed following the same scheme as described. However, in contrast to \cite{2016MNRAS.456.3866C,2017MNRAS.472.4900C}, who employed the formulation by \cite{2015A&A...582L...2S}, we have simply inhibited the BPs by halting the enlargement of the convective core whenever it would result in an increase of the central He abundance. In both cases (BP and non-BP sequences), and in agreement with \cite{2016MNRAS.456.3866C,2017MNRAS.472.4900C}, we find that 
our sequences develop a large partially mixed region with a stepped composition profile 
around the convective core. We mention that in all of our calculations, the Schwarzschild criterion for convection is used. In Table~\ref{tabla1} we list some relevant quantities of our sequences. 

Fig. \ref{fig:hecore} (see also Table ~\ref{tabla1}) clearly shows that the occurrence of  BPs extends the lifetime of CHeB phase as a result of the larger amount of He burned during this phase, as compared with the situation in which BP are suppressed. This shortens the lifetime of the following early-asymptotic giant branch (AGB) phase because of the less amount of available He that is left for this phase, as shown in  the bottom panel of Fig. \ref{fig:hecore}. This figure depicts the results for our  1.0 $M_{\odot}$ evolutionary sequence and are in line with the findings of \cite{2016MNRAS.456.3866C,2017MNRAS.472.4900C}.  The less amount of He that is left for the AGB phase as a result of the occurrence of BPs has no appreciable impact on the emerging WD. In particular, only minor differences are expected in the final H content and stellar mass of the WD, and in the third dredge-up episodes (as reflected by the final surface $^{12}$C to $^{16}$O ratio) that take place during the thermally pulsing AGB phase of WD progenitor. However, as we will see, the chemical structure of the core is substantially affected by the occurrence of BPs during the CHeB phase. This behavior is essentially the same also for the $2.5 M_{\sun}$ evolutionary sequence studied in this work.

\section{Chemical profiles}
\label{chemical_profiles}

\begin{figure}
\begin{center}
\includegraphics[clip,width=1.0\columnwidth]{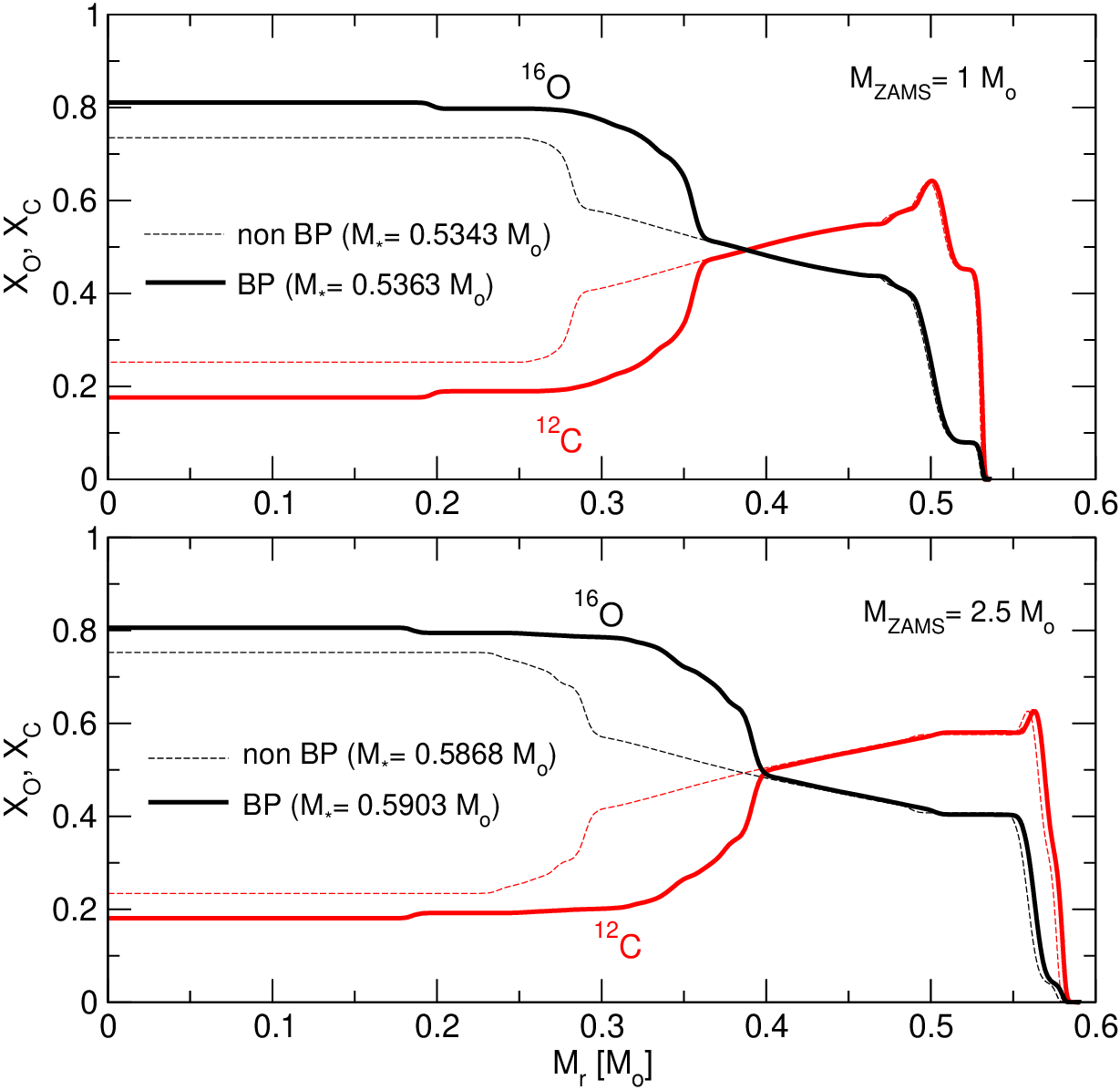}
        \caption{Chemical profiles of $^{16}$O (black) and $^{12}$C (red) in terms of the 
        mass coordinate (in solar mass units), corresponding to template WD models with $T_{\rm eff}= 12\,400$ K. Thin dashed (thick solid) lines correspond to the case
        in which BPs during the CHe-B evolution of the WD progenitor have been 
        suppressed (allowed to occur). Upper panel corresponds to 
        $M_{\star}= 0.5343 M_{\odot}$ (non-BP case) and $M_{\star}= 0.5363 M_{\odot}$ (BP case), whereas lower panel corresponds to $M_{\star}= 0.5868 M_{\odot}$ (non-BP case) and $M_{\star}= 0.5903 M_{\odot}$ (BP case).} 
        \label{fig:profile_Mr}
\end{center}
\end{figure}

We show in the upper panel of Fig. \ref{fig:profile_Mr} the $^{16}$O and $^{12}$C chemical profiles (fractional mass abundances) as a function of the mass coordinate (in units of solar masses), for two template WD models at $T_{\rm eff}= 12\,400$ K,  extracted from the $1.0 M_{\odot}$  WD progenitor evolution for the BP and non-BP cases (thick and thin lines, respectively).
The model corresponding to the non-BP case has a stellar mass of $M_{\star}= 0.5343 M_{\odot}$, and the model of the BP case is characterized by $M_{\star}= 0.5363 M_{\odot}$. The model that has experienced BPs exhibits a higher central abundance of $^{16}$O than the model that has not experienced BPs in the previous evolutionary history ($X_{\rm ^{16}O}= 0.808$ versus $X_{\rm ^{16}O}= 0.724$). A second relevant feature that can be seen in the figure is the larger size of the core for the BP case. Indeed, the total $^{16}$O content ($M_{\rm ^{16}O}$) in the BP case is $0.66 M_{\star}$, being for the non-BP case of $0.60 M_{\star}$. In the lower panel, we show the chemical profiles for the case of two template models extracted from the $2.5 M_{\odot}$  WD progenitor evolution for the BP and non-BP cases (thick and thin lines, respectively).
The results for these more massive models are qualitatively similar to those described before for the less massive model, that is, in the case in which BPs are allowed, the core is larger and the central abundance of $^{16}$O is higher than for the case in which the BPs are suppressed. 

These results were expected, and are in qualitative agreement with previous works \citep[e.g.,][]{2016MNRAS.456.3866C,2017MNRAS.472.4900C}.  
In particular, we obtain results in line also with the computations of \cite{2022FrASS...9.9045G}, who obtain more massive WD cores with larger central abundances of  $^{16}$O  when taking into account the occurrence of BPs. Nevertheless, we cannot help but notice that we are not able to obtain so high central abundance of $^{16}$O as they derive ($X_{\rm ^{16}O}= 0.86$) when considering BPs. 
Our results considering the occurrence of BPs neither explain the high central $^{16}$O abundance nor the huge $^{16}$O content of the core  asteroseismologically derived by  \cite{2018Natur.554...73G} for the DBV star KIC~08626021, of $X_{\rm ^{16}O}= 0.86$ and $M_{\rm ^{16}O}= 0.78 M_{\star}$, respectively.

\begin{figure}
\begin{center}
\includegraphics[clip,width=1.0\columnwidth]{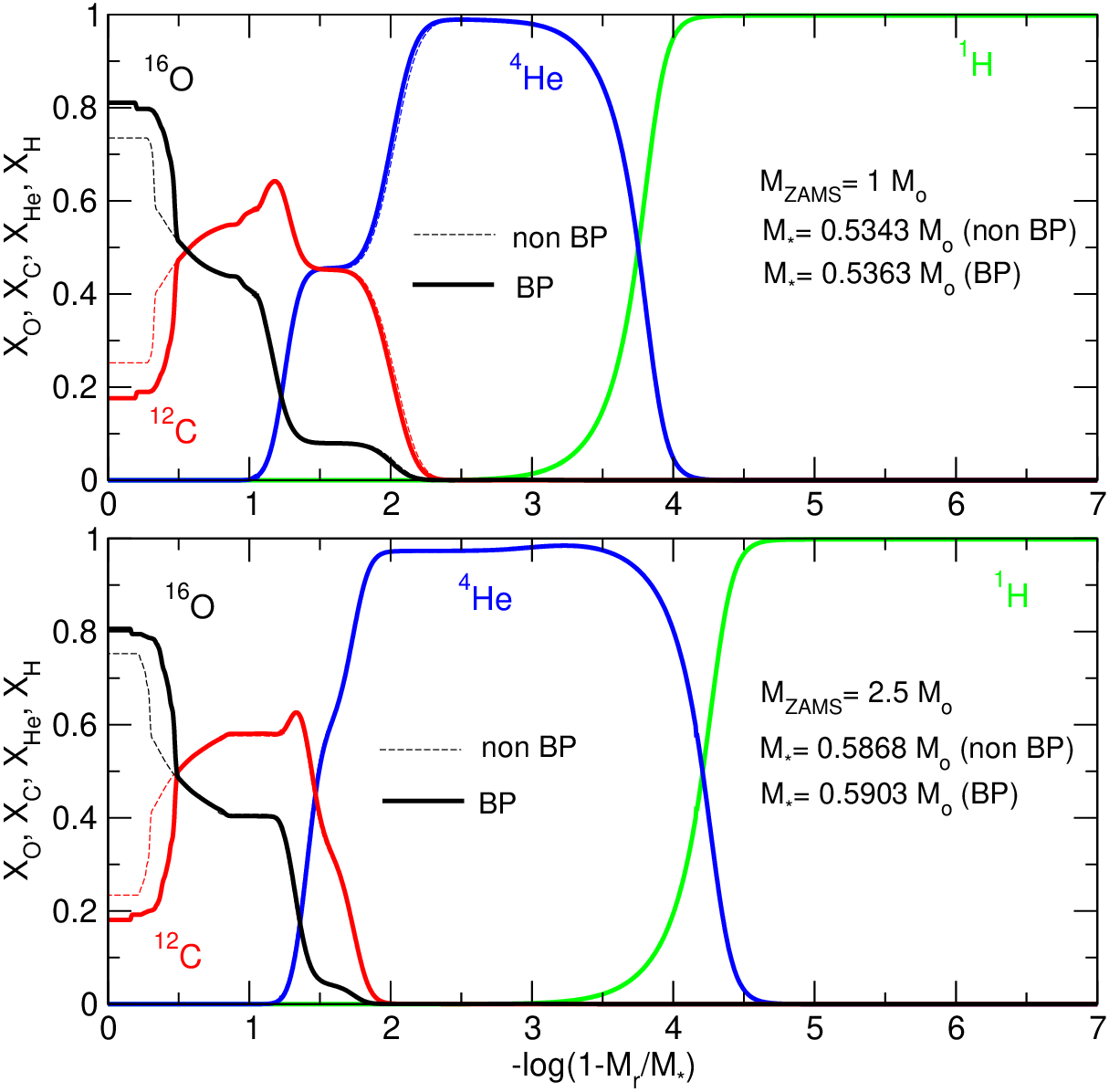}
        \caption{Chemical profiles of $^{16}$O (black), $^{12}$C (red), $^{4}$He (blue), and $^{1}$H (green) in terms of outer mass fraction coordinate, corresponding to the same template WD models with $T_{\rm eff}= 12\,400$ K shown in Fig. \ref{fig:profile_Mr}. Again, thin dashed lines correspond to the non-BP case, while solid thick lines correspond to the BP case.} 
        \label{fig:profile_outer_mass}
\end{center}
\end{figure}

It should be noted that the evolutionary calculations of \cite{2022FrASS...9.9045G} do not include the AGB phase, but are restricted to producing Extreme Horizontal Branch (EHB) models and evolving them through the He-core and He-shell burning phases, and then let them contract and cool on the WD sequence.
Since our simulations include the complete evolution from the ZAMS through the AGB phase to the WD stage, our analysis is able to  account for all the processes that ultimately shape the WD internal
chemical stratification from the center to the surface. We depict in Fig. \ref{fig:profile_outer_mass} the complete chemical structure of the species $^{16}$O, $^{12}$C, $^{4}$He, and $^{1}$H of the same template WDs models presented in Fig. \ref{fig:profile_Mr}, as a function of the logarithm of the outer mass fraction. The imprints of the occurrence of thermal pulses during progenitor evolution manifest themselves as the presence of the intershell rich in helium, carbon and, oxygen, as illustrated in the upper panel of Fig. \ref{fig:profile_Mr}. This intershell is not present in the WD models resulting from the more massive progenitor (see bottom panel) because in this case element diffusion turns out to be much more efficient in shaping the final chemical structure of the intershell by the time the ZZ domain is reached.  Apart from the differences in the size of the core and in the central abundance of $^{16}$O already displayed in Fig. \ref{fig:profile_Mr}, this plot shows that there are no appreciable differences in the chemical profiles in other parts of the models, resulting from the occurrence of BPs. The differences in the core chemical structure have consequences on the period spectrum of the WDs. We focus on this issue in the next section.

\section{Pulsations}
\label{pulsation}

We assess the impact of BPs during the CHeB phase on the pulsation spectrum of WDs by comparing the $g$ mode period spectrum of models that were computed in the non-BP case and models constructed in the BP case. The pulsation modes of our DA WD models have been computed with the adiabatic version of the {\tt  LP-PUL} pulsation code described in \citet{2006A&A...454..863C}.  The squared Brunt-V\"ais\"al\"a  frequency ($N$, the critical frequency of nonradial $g$-mode pulsations) is computed as in \cite{1990ApJS...72..335T}, according to the following expression:

\begin{equation}
N^2= \frac{g^2 \rho}{P}\frac{\chi_{\rm T}}{\chi_{\rho}}
\left[\nabla_{\rm ad}- \nabla + B\right],
\label{bv}
\end{equation}

\noindent where $g$, $\rho$, $P$, $\nabla_{\rm ad}$ and $\nabla$ are 
the acceleration of gravity, density, pressure, adiabatic temperature 
gradient and actual temperature gradient, respectively. The 
compressibilities $\chi_{\rho}$ and $\chi_{\rm T}$ are defined as:

\begin{equation}
\chi_{\rho}= \left(\frac{d\ln P}{d\ln \rho}\right)_{{\rm T}, \{\rm X_i\}}\ \ \
\chi_{\rm T}= \left(\frac{d\ln P}{d\ln T}\right)_{\rho, \{\rm X_i\}}.
\label{compre}
\end{equation}

Finally, the Ledoux term $B$ is computed as \citep{1990ApJS...72..335T}:

\begin{equation}
B= -\frac{1}{\chi_{\rm T}} \sum_1^{M-1} \chi_{\rm X_i} \frac{d\ln X_i}{d\ln P}, 
\label{BLedoux}
\end{equation}

\noindent where

\begin{equation}
\chi_{\rm X_i}= \left(\frac{d\ln P}{d\ln X_i}\right)_{\rho, {\rm T}, 
\{\rm X_{j \neq i}\}}.
\label{compre_chi}
\end{equation}

The  computation of  the Ledoux term includes the effects of an 
arbitrary number of chemical species that vary in abundance in the 
transition regions. For completeness, we also calculate the Lamb frequency 
($L_{\ell}$, the critical frequency of nonradial $p$-mode pulsations) 
according to the expression:

\begin{equation}
L_{\ell}^2= \ell(\ell+1)\frac{c_{\rm s}^2}{r^2}
\label{compre_chi}
\end{equation}

\noindent where $c_{\rm s}$ is the local velocity of sound.

\begin{figure}
\begin{center}
\includegraphics[clip,width=1.0\columnwidth]{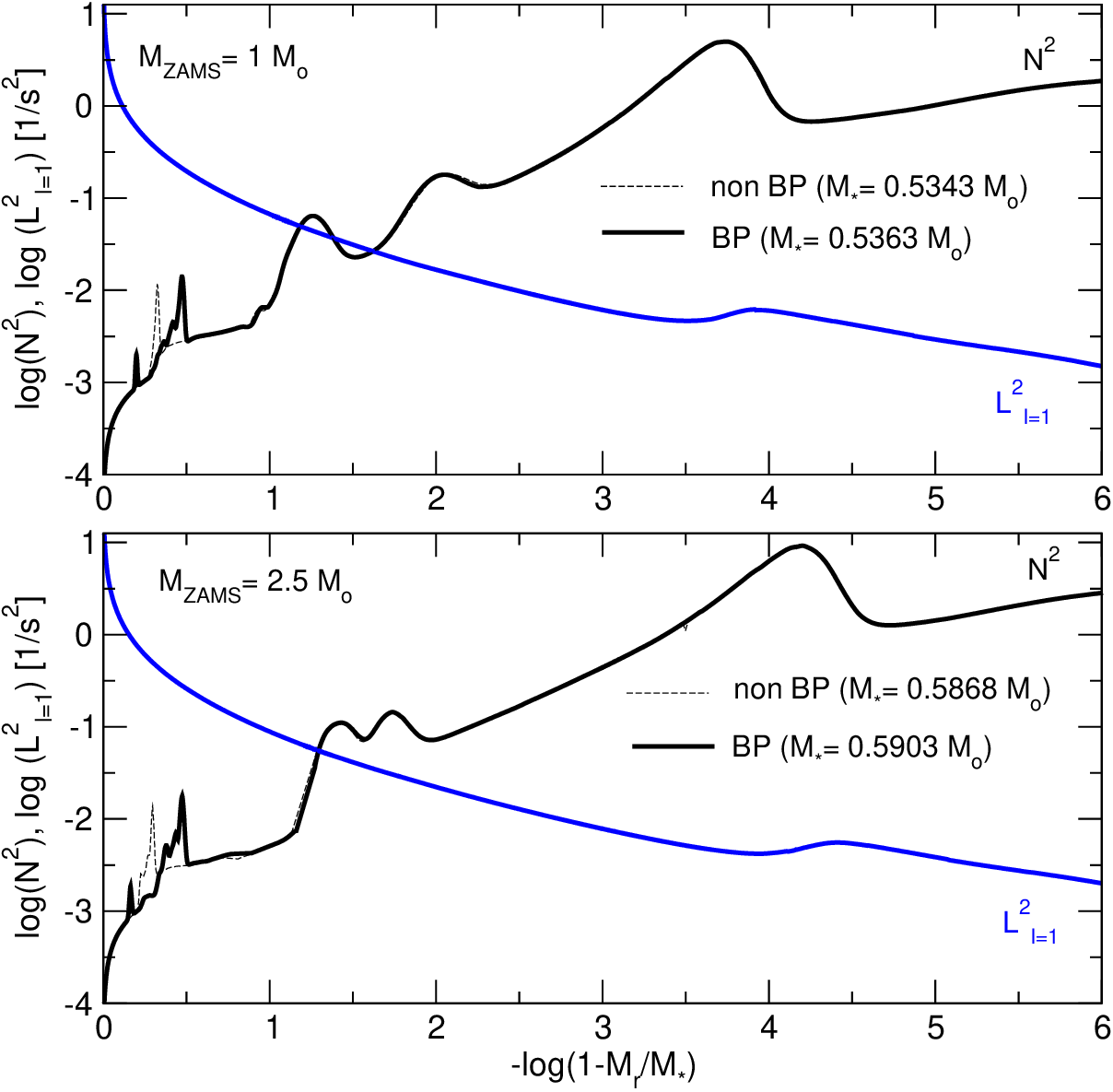}
        \caption{Logarithm of the squared Brunt-V\"ais\"al\"a (black) and Lamb (blue) critical frequencies for $\ell= 1$ modes in terms of the outer mass fraction coordinate, corresponding to the same template WD models with $T_{\rm eff}= 12\,400$ K depicted in Figs. \ref{fig:profile_Mr} and \ref{fig:profile_outer_mass}. Thin dashed (thick solid) lines correspond to the non-BP case
        (BP case).} 
        \label{fig:FBV}
\end{center}
\end{figure}

The run of the logarithm of the squared Brunt-V\"ais\"al\"a and Lamb ($\ell= 1$) frequencies for the same template models displayed in Figs. \ref{fig:profile_Mr} and \ref{fig:profile_outer_mass} are shown in the two panels of Fig. \ref{fig:FBV}. As can be seen in the figure, all the the bumps of the Brunt-V\"ais\"al\"a frequency are located at the same places within the models, except in the case of the C/O chemical transition at the core. Indeed, the main peak of $N^2$ at the core of the BP case models is located further out than in the non-BP case models. Additionally, in the BP case models, there exists an additional peak, located in deeper regions of the core. This last peak is due to the small step exhibited by the $^{16}$O and $^{12}$C profiles at $-\log(1-M_r/M_{\star})\sim 0.2$ (see Fig. \ref{fig:profile_outer_mass}).

\begin{figure}
\begin{center}
\includegraphics[clip,width=1.0\columnwidth]{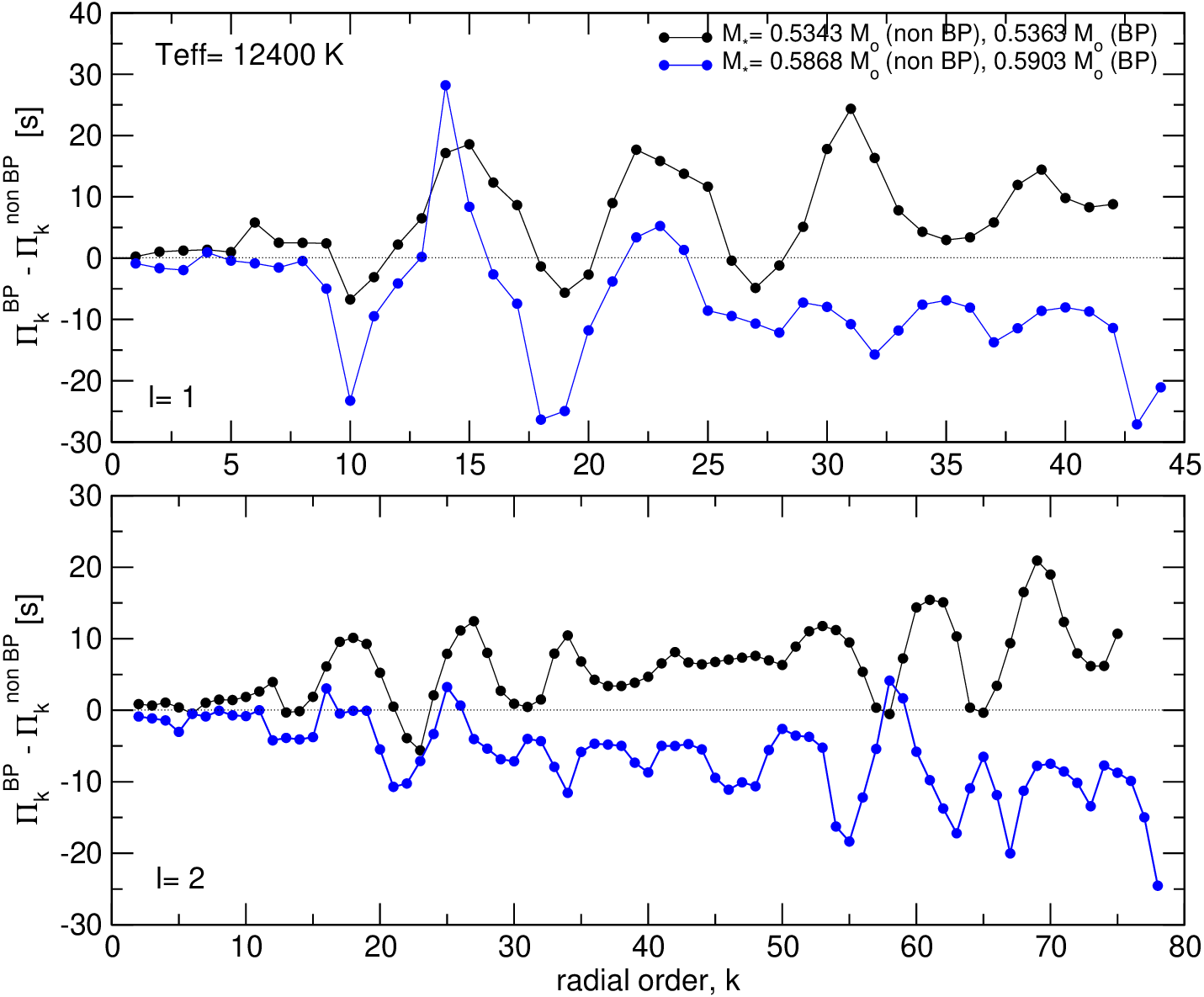}
        \caption{Period differences (with fixed radial order $k$) in terms of $k$ between the BP and non-BP template models with $T_{\rm eff}= 12\,400$ K and stellar 
        masses $M_{\star} \sim 0.53 M_{\sun}$ ($M_{\rm ZAMS}= 1.0 M_{\sun}$)
        (black) and $M_{\star}\sim 0.59 M_{\sun}$ ($M_{\rm ZAMS}= 2.5 M_{\sun}$) (blue), corresponding to $\ell= 1$ (upper panel) and $\ell= 2$ (lower panel) modes.} 
        \label{fig:delta_P}
\end{center}
\end{figure}

The differences in the spatial location and number of the bumps in the profile of $N^2$ at the core of the template WD models 
have sizeable consequences on the pulsation periods and period spacings of the $g$ modes. For the case of individual periods, this is evident in Fig. \ref{fig:delta_P}, in which we have plotted the difference between the dipole ($\ell= 1$) and quadrupole ($\ell= 2$) periods of $g$ modes calculated in the BP case ($\Pi_k^{\rm BP}$) and the periods calculated in the non-BP case ($\Pi_k^{\rm non-BP}$) in terms of the radial order $k$, for the same template WD models considered in the previous figures. 
As can be seen, the absolute value of the difference in the periods can reach up to $\sim 30$ s for modes with periods in the range $100-2000$ s, the period interval in which the periods observed in ZZ Ceti stars generally fall. The differences in the periods come almost exclusively from the different chemical structures of the core of the WDs depending on the case (BP and non-BP). The tiny difference in mass  ($|\Delta M_{\star}|= |M_{\star}^{\rm BP}-M_{\star}^{\rm non-BP}| \lesssim 0.0035  M_{\sun}$) between the couples of template models has a negligible impact on the periods.

\begin{figure}
\begin{center}
\includegraphics[clip,width=1.0\columnwidth]{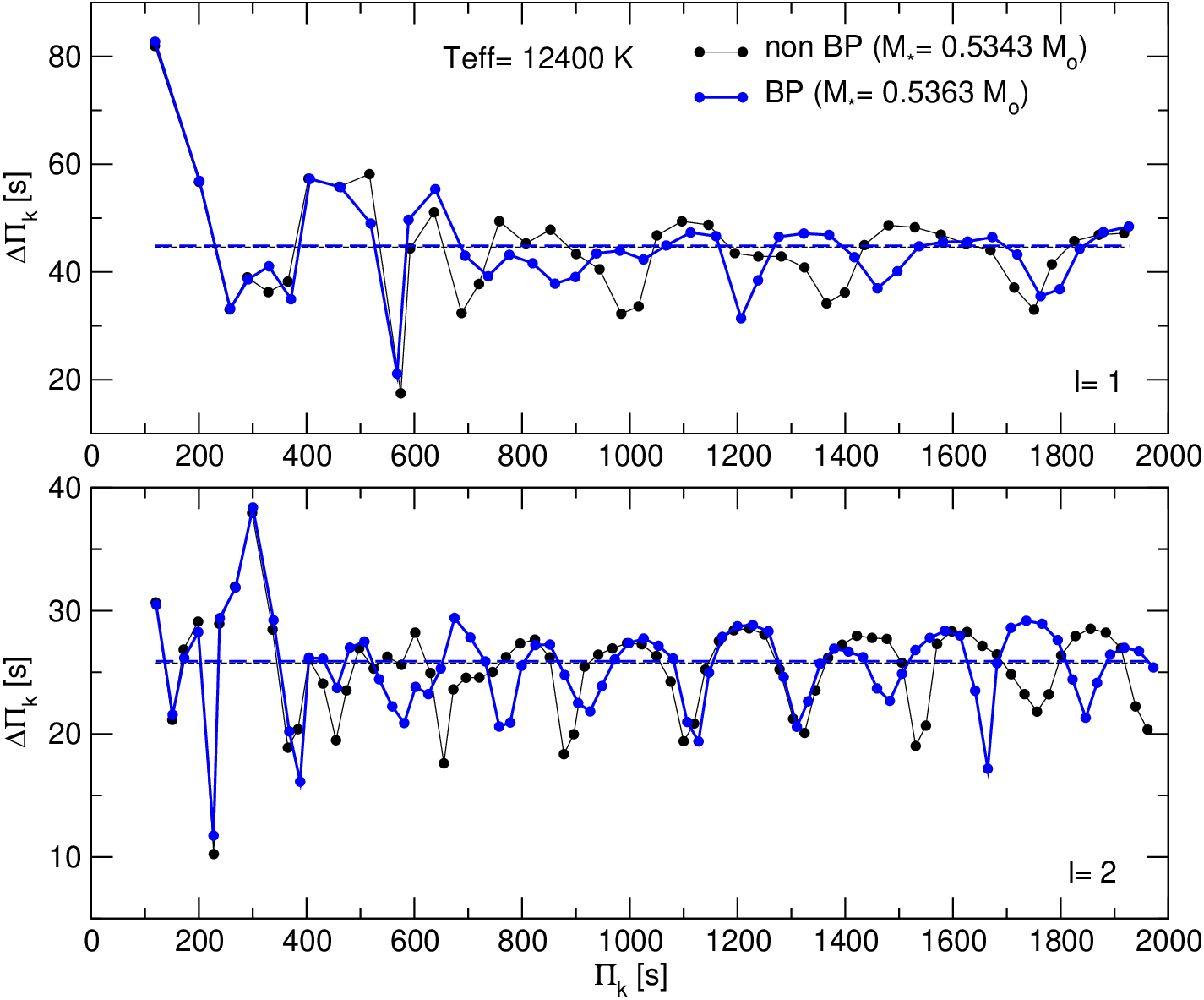}
        \caption{Forward period spacings versus periods for $\ell= 1$ (upper panel) and $\ell= 2$ (lower panel) $g$ modes, corresponding to the DA WD template models with $T_{\rm eff}= 12\,400$ K representative of ZZ Cetis stars. Black (blue) dots connected with black thin (blue thick) lines correspond to the non-BP (BP) case for a ZZ Ceti model with $M_{\star}= 0.5343 M_{\sun}$ ($M_{\star}= 0.5343 M_{\sun}$).}
        \label{fig:DELP_P}
\end{center}
\end{figure}

\begin{figure}
\begin{center}
\includegraphics[clip,width=1.0\columnwidth]{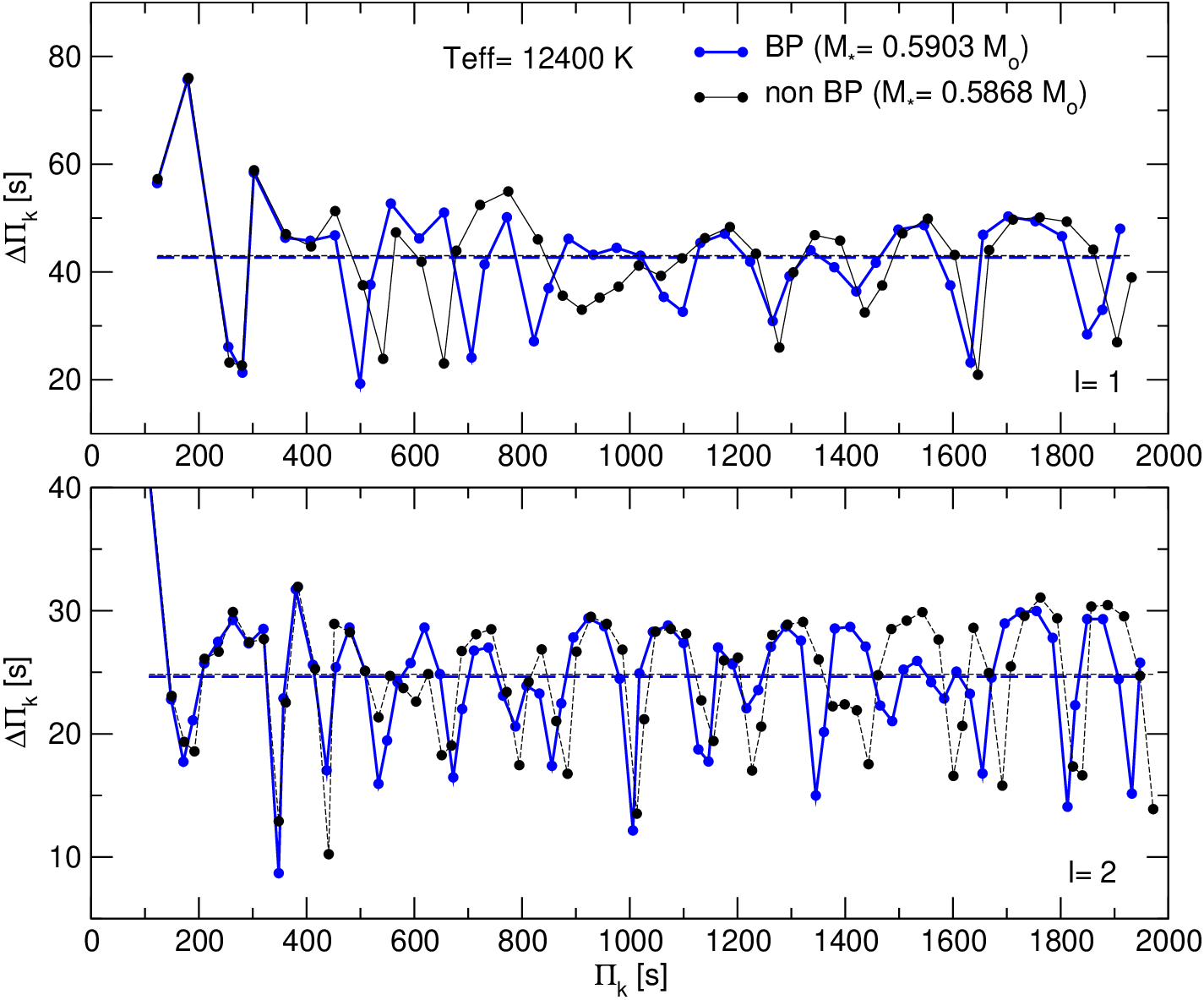}
        \caption{Same as Fig. \ref{fig:DELP_P}, but for ZZ Ceti models with $M_{\star}= 0.5868 M_{\sun}$ (non BP case) and $M_{\star}= 0.5903 M_{\sun}$ (BP case).} 
        \label{fig:DELP_P_2}
\end{center}
\end{figure}

In closing, we depict in Figs. \ref{fig:DELP_P} and Figs. \ref{fig:DELP_P_2} the forward period spacing, defined as $\Delta \Pi_k\equiv \Pi_{k+1} - \Pi_k$, in terms of periods ($\Pi_k$), for dipole (upper panel) and quarupole (lower panel) modes corresponding to the case in which BPs have been considered (blue symbols and thick lines) and the situation in which BPs have been suppressed (black symbols and thin lines) during the CHeB phase. As it can be seen, the general appearance of the period-spacing distribution is similar for the BP and the non-BP cases, but there are substantial quantitative differences for periods longer than $\sim 400$ s. The asymptotic period spacing (horizontal dashed lines) is slightly different between both cases, due to the small difference of stellar mass of the models of the non-BP and BP cases. We conclude that the mean period spacing of pulsating DA WDs is insensitive to the occurrence or not of BPs. However, seismological period-to-period fits of ZZ Ceti stars based on evolutionary models generated considering and neglecting the occurrence of BPs during core-He burning, could help shed some light on the occurrence of BPs in Nature.

\section{Summary and conclusions}
\label{conclusions}

In this work, we have revisited the issue of BPs, which consist of mixing events that can occur at the end of the CHeB phase during the evolution of low- and intermediate-mass stars \citep[][]{1973ASSL...36..221S,1985ApJ...296..204C}. The occurrence or not of BPs is expected to have influence on the evolution of WDs,  in
particular on their core chemical structure, which can be  probed through asteroseismology. 
Interestingly enough, recent studies of pulsating WDs \citep{2018Natur.554...73G, 2021arXiv210615701G, 2021arXiv210703797C} point to asteroseismological models characterized by central $^{16}$O abundances and core sizes significantly larger than standard WD formation theory indicates. This has given rise to the belief that the cores of WDs in general should be larger and more $^{16}$O-rich than previously believed, suggesting that some piece of physics in the formation of WDs has been missing until now \citep {2018Natur.554...73G}. In particular, \cite{2022FrASS...9.9045G} have claimed that a possible explanation for these anomalous properties of the WD cores could be at the root of the BP episodes during CHeB phase. 

We have carried out simulations describing the complete evolution of low-mass star progenitors with $Z=0.01$ evolved from the ZAMS, through the CHeB phase, to the thermal pulses on the AGB, and finally to the domain of the ZZ Ceti stars at advanced stages of WD cooling. We have considered two initial masses at the ZAMS ($M_{\rm ZAMS}/M_{\sun}= 1.0, 2.5$) taking into account and neglecting the occurrence of BPs during the CHeB phase. We arrive at two important results:

\begin{itemize}

\item We confirm previous results \citep[e.g.,][]{2016MNRAS.456.3866C,2017MNRAS.472.4900C} that the occurrence of BPs induces the formation of more massive and $^{16}$O-rich cores compared to the case in which BPs have been ignored. At this point, however, we cannot help but notice that the occurrence of BPs is not at all sufficient to explain the excessively large sizes of the WD cores and the anomalously high central $^{16}$O abundances predicted by recent asteroseismological studies \citep{2018Natur.554...73G, 2021arXiv210615701G, 2021arXiv210703797C}. However, in line with the finding of these authors,  a recent self-consistent implementation of convective penetration during CHeB phase \citep{2023arXiv231208315J} predicts the formation of larger, $^{16}$O-rich WD cores, which would naturally produce more massive remnant C/O cores at the end of He burning. The consequences of convective penetration on the core size and central oxygen abundance of WDs is a topic that deserves to be explored in the future.

\item Our pulsational analysis indicates that the occurrence of BPs can lead to $g$-mode periods of ZZ Ceti stars that can differ by up to $\sim 30$ seconds (excess or defect) compared to the situation in which BPs do not occur during the evolution of the progenitors.  It is not surprising that the presence or absence of BPs during the CHeB phase has a significant effect on $g$ mode periods, since  Fig. \ref{fig:FBV} shows that there is clear difference in $N^2$ between the BP and non-BP cases in the
C/O chemical transition at the core. The mean period spacing of pulsating DA WDs is insensitive to the occurrence or not of BPs, while the forward period spacing shows appreciable differences for periods longer than $\sim 400$ s.

\end{itemize}

We conclude that future seismological period-to-period fits of DAV stars based on evolutionary models generated considering and neglecting the occurrence of BPs during core-He burning, could
help shed some light on the occurrence of BPs in Nature, and  comparable outcome could probably be attained through the analysis of DBV stars.

\section*{Acknowledgments}  
  
We  wish  to  thank  the  suggestions  and comments of an anonymous referee that improved the original version of this work. 
Part of this work was supported by AGENCIA through the Programa de Modernizaci\'on Tecnol\'ogica BID 1728/OC-AR, and by the PIP 112-200801-00940 grant from CONICET. This  research has  made use of  NASA Astrophysics Data System.

\bibliographystyle{aasjournal}
\bibliography{breathing}

\begin{thebibliography}{}
\expandafter\ifx\csname natexlab\endcsname\relax\def\natexlab#1{#1}\fi
\providecommand{\url}[1]{\href{#1}{#1}}
\providecommand{\dodoi}[1]{doi:~\href{http://doi.org/#1}{\nolinkurl{#1}}}
\providecommand{\doeprint}[1]{\href{http://ascl.net/#1}{\nolinkurl{http://ascl.net/#1}}}
\providecommand{\doarXiv}[1]{\href{https://arxiv.org/abs/#1}{\nolinkurl{https://arxiv.org/abs/#1}}}

\bibitem[{{Althaus} {et~al.}(2015){Althaus}, {Camisassa}, {Miller Bertolami},
  {C{\'o}rsico}, \& {Garc{\'{\i}}a-Berro}}]{2015A&A...576A...9A}
{Althaus}, L.~G., {Camisassa}, M.~E., {Miller Bertolami}, M.~M., {C{\'o}rsico},
  A.~H., \& {Garc{\'{\i}}a-Berro}, E. 2015, \aap, 576, A9,
  \dodoi{10.1051/0004-6361/201424922}

\bibitem[{{Althaus} \& {C{\'o}rsico}(2022)}]{2022A&A...663A.167A}
{Althaus}, L.~G., \& {C{\'o}rsico}, A.~H. 2022, \aap, 663, A167,
  \dodoi{10.1051/0004-6361/202243943}

\bibitem[{{Althaus} {et~al.}(2010{\natexlab{a}}){Althaus}, {C{\'o}rsico},
  {Bischoff-Kim}, {Romero}, {Renedo}, {Garc{\'{\i}}a-Berro}, \& {Miller
  Bertolami}}]{2010ApJ...717..897A}
{Althaus}, L.~G., {C{\'o}rsico}, A.~H., {Bischoff-Kim}, A., {et~al.}
  2010{\natexlab{a}}, \apj, 717, 897, \dodoi{10.1088/0004-637X/717/2/897}

\bibitem[{{Althaus} {et~al.}(2010{\natexlab{b}}){Althaus}, {C{\'o}rsico},
  {Isern}, \& {Garc{\'{\i}}a-Berro}}]{2010A&ARv..18..471A}
{Althaus}, L.~G., {C{\'o}rsico}, A.~H., {Isern}, J., \& {Garc{\'{\i}}a-Berro},
  E. 2010{\natexlab{b}}, \aapr, 18, 471, \dodoi{10.1007/s00159-010-0033-1}

\bibitem[{{Althaus} {et~al.}(2005){Althaus}, {Serenelli}, {Panei},
  {C{\'o}rsico}, {Garc{\'{\i}}a-Berro}, \&
  {Sc{\'o}ccola}}]{2005A&A...435..631A}
{Althaus}, L.~G., {Serenelli}, A.~M., {Panei}, J.~A., {et~al.} 2005, \aap, 435,
  631, \dodoi{10.1051/0004-6361:20041965}

\bibitem[{{Bell}(2022)}]{2022RNAAS...6..244B}
{Bell}, K.~J. 2022, Research Notes of the American Astronomical Society, 6,
  244, \dodoi{10.3847/2515-5172/aca3ad}

\bibitem[{{Castellani} {et~al.}(1985){Castellani}, {Chieffi}, {Tornambe}, \&
  {Pulone}}]{1985ApJ...296..204C}
{Castellani}, V., {Chieffi}, A., {Tornambe}, A., \& {Pulone}, L. 1985, \apj,
  296, 204, \dodoi{10.1086/163437}

\bibitem[{{Charpinet} {et~al.}(2019){Charpinet}, {Brassard}, {Giammichele}, \&
  {Fontaine}}]{2019A&A...628L...2C}
{Charpinet}, S., {Brassard}, P., {Giammichele}, N., \& {Fontaine}, G. 2019,
  \aap, 628, L2, \dodoi{10.1051/0004-6361/201935823}

\bibitem[{{Charpinet} {et~al.}(2021){Charpinet}, {Giammichele}, {Brassard},
  {Fontaine}, {Bergeron}, {Zong}, {Van Grootel}, \&
  {Baran}}]{2021arXiv210703797C}
{Charpinet}, S., {Giammichele}, N., {Brassard}, P., {et~al.} 2021, arXiv
  e-prints, arXiv:2107.03797, \dodoi{10.48550/arXiv.2107.03797}

\bibitem[{{Constantino} {et~al.}(2017){Constantino}, {Campbell}, \&
  {Lattanzio}}]{2017MNRAS.472.4900C}
{Constantino}, T., {Campbell}, S.~W., \& {Lattanzio}, J.~C. 2017, \mnras, 472,
  4900, \dodoi{10.1093/mnras/stx2321}

\bibitem[{{Constantino} {et~al.}(2016){Constantino}, {Campbell}, {Lattanzio},
  \& {van Duijneveldt}}]{2016MNRAS.456.3866C}
{Constantino}, T., {Campbell}, S.~W., {Lattanzio}, J.~C., \& {van Duijneveldt},
  A. 2016, \mnras, 456, 3866, \dodoi{10.1093/mnras/stv2939}

\bibitem[{{C{\'o}rsico} \& {Althaus}(2006)}]{2006A&A...454..863C}
{C{\'o}rsico}, A.~H., \& {Althaus}, L.~G. 2006, \aap, 454, 863,
  \dodoi{10.1051/0004-6361:20054199}

\bibitem[{{C{\'o}rsico} {et~al.}(2019){C{\'o}rsico}, {Althaus}, {Miller
  Bertolami}, \& {Kepler}}]{2019A&ARv..27....7C}
{C{\'o}rsico}, A.~H., {Althaus}, L.~G., {Miller Bertolami}, M.~M., \& {Kepler},
  S.~O. 2019, \aapr, 27, 7, \dodoi{10.1007/s00159-019-0118-4}

\bibitem[{{De Ger{\'o}nimo} {et~al.}(2019){De Ger{\'o}nimo}, {Battich}, {Miller
  Bertolami}, {Althaus}, \& {C{\'o}rsico}}]{2019A&A...630A.100D}
{De Ger{\'o}nimo}, F.~C., {Battich}, T., {Miller Bertolami}, M.~M., {Althaus},
  L.~G., \& {C{\'o}rsico}, A.~H. 2019, \aap, 630, A100,
  \dodoi{10.1051/0004-6361/201834988}

\bibitem[{{Fontaine} \& {Brassard}(2008)}]{2008PASP..120.1043F}
{Fontaine}, G., \& {Brassard}, P. 2008, PASP, 120, 1043, \dodoi{10.1086/592788}

\bibitem[{{Giammichele} {et~al.}(2022){Giammichele}, {Charpinet}, \&
  {Brassard}}]{2022FrASS...9.9045G}
{Giammichele}, N., {Charpinet}, S., \& {Brassard}, P. 2022, Frontiers in
  Astronomy and Space Sciences, 9, 879045, \dodoi{10.3389/fspas.2022.879045}

\bibitem[{{Giammichele} {et~al.}(2021){Giammichele}, {Charpinet}, {Fontaine},
  {Brassard}, {Bergeron}, {Reindl}, \& {Baran}}]{2021arXiv210615701G}
{Giammichele}, N., {Charpinet}, S., {Fontaine}, G., {et~al.} 2021, arXiv
  e-prints, arXiv:2106.15701, \dodoi{10.48550/arXiv.2106.15701}

\bibitem[{{Giammichele} {et~al.}(2018){Giammichele}, {Charpinet}, {Fontaine},
  {Brassard}, {Green}, {Van Grootel}, {Bergeron}, {Zong}, \&
  {Dupret}}]{2018Natur.554...73G}
---. 2018, \nat, 554, 73, \dodoi{10.1038/nature25136}

\bibitem[{{Johnston} {et~al.}(2023){Johnston}, {Michielsen}, {Anders}, {Renzo},
  {Cantiello}, {Marchant}, {Goldberg}, {Townsend}, {Sabhahit}, \&
  {Jermyn}}]{2023arXiv231208315J}
{Johnston}, C., {Michielsen}, M., {Anders}, E.~H., {et~al.} 2023, arXiv
  e-prints, arXiv:2312.08315.
\newblock \doarXiv{2312.08315}

\bibitem[{{Kippenhahn} {et~al.}(2013){Kippenhahn}, {Weigert}, \&
  {Weiss}}]{2013sse..book.....K}
{Kippenhahn}, R., {Weigert}, A., \& {Weiss}, A. 2013, {Stellar Structure and
  Evolution}, \dodoi{10.1007/978-3-642-30304-3}

\bibitem[{{Miller Bertolami}(2016)}]{2016A&A...588A..25M}
{Miller Bertolami}, M.~M. 2016, \aap, 588, A25,
  \dodoi{10.1051/0004-6361/201526577}

\bibitem[{{Renedo} {et~al.}(2010){Renedo}, {Althaus}, {Miller Bertolami},
  {Romero}, {C{\'o}rsico}, {Rohrmann}, \&
  {Garc{\'{\i}}a-Berro}}]{2010ApJ...717..183R}
{Renedo}, I., {Althaus}, L.~G., {Miller Bertolami}, M.~M., {et~al.} 2010, \apj,
  717, 183, \dodoi{10.1088/0004-637X/717/1/183}

\bibitem[{{Salaris} {et~al.}(2013){Salaris}, {Althaus}, \&
  {Garc{\'\i}a-Berro}}]{2013A&A...555A..96S}
{Salaris}, M., {Althaus}, L.~G., \& {Garc{\'\i}a-Berro}, E. 2013, \aap, 555,
  A96, \dodoi{10.1051/0004-6361/201220622}

\bibitem[{{Salaris} {et~al.}(2010){Salaris}, {Cassisi}, {Pietrinferni},
  {Kowalski}, \& {Isern}}]{2010ApJ...716.1241S}
{Salaris}, M., {Cassisi}, S., {Pietrinferni}, A., {Kowalski}, P.~M., \&
  {Isern}, J. 2010, \apj, 716, 1241, \dodoi{10.1088/0004-637X/716/2/1241}

\bibitem[{{Spruit}(2015)}]{2015A&A...582L...2S}
{Spruit}, H.~C. 2015, \aap, 582, L2, \dodoi{10.1051/0004-6361/201527171}

\bibitem[{{Sweigart} \& {Demarque}(1973)}]{1973ASSL...36..221S}
{Sweigart}, A.~V., \& {Demarque}, P. 1973, in Astrophysics and Space Science
  Library, Vol.~36, IAU Colloq. 21: Variable Stars in Globular Clusters and in
  Related Systems, ed. J.~D. {Fernie}, 221,
  \dodoi{10.1007/978-94-010-2590-4_32}

\bibitem[{{Tassoul} {et~al.}(1990){Tassoul}, {Fontaine}, \&
  {Winget}}]{1990ApJS...72..335T}
{Tassoul}, M., {Fontaine}, G., \& {Winget}, D.~E. 1990, ApJs, 72, 335,
  \dodoi{10.1086/191420}

\bibitem[{{Timmes} {et~al.}(2018){Timmes}, {Townsend}, {Bauer}, {Thoul},
  {Fields}, \& {Wolf}}]{2018ApJ...867L..30T}
{Timmes}, F.~X., {Townsend}, R.~H.~D., {Bauer}, E.~B., {et~al.} 2018, \apjl,
  867, L30, \dodoi{10.3847/2041-8213/aae70f}

\bibitem[{{Winget} \& {Kepler}(2008)}]{2008ARA&A..46..157W}
{Winget}, D.~E., \& {Kepler}, S.~O. 2008, \araa, 46, 157,
  \dodoi{10.1146/annurev.astro.46.060407.145250}

\bibitem[{{Woosley} \& {Heger}(2015)}]{2015ApJ...810...34W}
{Woosley}, S.~E., \& {Heger}, A. 2015, \apj, 810, 34,
  \dodoi{10.1088/0004-637X/810/1/34}

\end{thebibliography}

\end{document}